\documentclass[sn-mathphys]{sn-jnl}

\usepackage{amsmath}
\usepackage{amssymb}
\usepackage{bm}
\usepackage{hyperref}

\jyear{2026}%

\theoremstyle{thmstyleone}%
\theoremstyle{thmstyletwo}%

\theoremstyle{thmstylethree}%

\begin{document}

\title[Fourier PINN adhesive contact mechanics]{Mass weighting algorithm optimizes Fourier-based
physics-informed neural network in adhesive contact mechanics}


\author*[1]{\fnm{Yunong} \sur{Zhou}}\email{yunong.zhou@yzu.edu.cn}

\author[1]{\fnm{Kaifeng} \sur{Huang}}

\author[1]{\fnm{Chaofan} \sur{Du}}

\author[2,3]{\fnm{Yang} \sur{Xu}}

\author*[4,5]{\fnm{Hengxu} \sur{Song}}\email{songhengxu@imech.ac.cn}

\affil[1]{\orgdiv{Department of Civil Engineering}, \orgname{Yangzhou University}, \city{Yangzhou},
\postcode{225127}, \state{Jiangsu}, \country{China}}

\affil[2]{\orgdiv{School of Mechanical Engineering}, \orgname{Hefei University of Technology},
\city{Hefei}, \postcode{230009}, \state{Anhui}, \country{China}}

\affil[3]{\orgname{Anhui Province Key Laboratory of Digital Design and Manufacturing},
\city{Hefei}, \postcode{230009}, \state{Anhui}, \country{China}}

\affil[4]{\orgdiv{LNM}, \orgname{Chinese Academy of Sciences},
\city{Beijing}, \postcode{100190}, \country{China}}

\affil[5]{\orgname{University of Chinese Academy of Sciences},
\city{Beijing}, \postcode{100049}, \country{China}}

\abstract{
Physics-informed neural networks (PINNs) for elastic contact mechanics suffer
from a spectral stiffness imbalance,that is, the elastic kernel grows linearly 
with wave number, causing short-wavelength modes to dominate gradient updates
and stall convergence of the macroscopic deformation.
We introduce a spectral preconditioning strategy that reweights loss
gradients with respect to displacements in Fourier space before back-propagation,
amplifying low wavenumber components through a mass weighting (MW) function while suppressing
sub-grid noise via a built-in low-pass filter.
Applied to adhesive line contact problems, the mass weighted PINN reaches
machine-zero residual loss within $400$ Adam iterations for specified benchmark,
whereas the reference benchmark stalls at three orders of magnitude higher loss.
The converged displacement and contact stress fields agree quantitatively with
Green's function molecular dynamics (GFMD) solutions for both
smooth Hertz contact at pressures spanning tension to compression and
rough surfaces with roughness covering several decades of wavelength.
The method operates directly on a uniform real-space grid, requires no
explicit Green's function integration or quadrature rules, and is formulated
entirely in terms of minimising a scalar energy function.
Extension to two-dimensional rough surfaces is direct, as both the Fourier
elastic energy and the spectral preconditioner depend only on the wave-number
magnitude.
}

\keywords{Contact mechanics, Adhesion, PINN, Mass weighting}



\maketitle

\section{Introduction}\label{sec:intro}

Contact between deformable bodies widely exists in both natural phenomena and
engineering applications.
For example, gecko adhesion relies on van der Waals contact interactions
between spatula-shaped setae and substrate surfaces~\cite{Gao2005MM,Zeng2009L,
Materzok2022S}.
In engineering practice, contact mechanics governs the performance and
reliability of tires gripping road surfaces~\cite{Persson2025JCP,Xu2025TL},
seals and gaskets preventing fluid leakage~\cite{Dapp2012PRL,Mueller2023PRL},
micro-electromechanical systems where stiction between contacting components
causes device failure~\cite{Li2024IJMS}, and nanoscale manufacturing processes
such as nano-imprint lithography and transfer printing~\cite{McClelland2005TL,
Carlson2012AFM}.
A unifying feature across these disparate systems is that the contact behavior,
including the real contact area, the stress distribution, the gap distribution,
and the adhesion strength, is governed by competing elastic deformation and
surface interaction forces at various length scales.

The challenge of predicting these emergent quantities has driven the development
of contact mechanics for decades.
Hertz provided the first rigorous solution for non-adhesive elastic
spheres~\cite{Hertz1882}.
Johnson, Kendall and Roberts (JKR)~\cite{Johnson1971RSPA} and Derjaguin,
Muller, and Toporov (DMT)~\cite{Derjaguin1975JCIS} incorporated surface energy
into the analysis, establishing two limiting regimes for adhesive contacts.
Maugis~\cite{Maugis1992JCIS} unified these limits through a Dugdale
cohesive-zone model, and the resulting Maugis-Tabor parameter provides
a standard framework for classifying adhesive contact
behavior~\cite{Ciavarella2019JMPS,Ciavarella2022JA}.
These classical solutions, however, apply exclusively to smooth, idealized
geometries and cannot capture the effects of multi-scale surface roughness.

Real engineering surfaces are rough across many length scales, and the
statistical properties of surface topography fundamentally dominates the
contact response.
Greenwood and Williamson (GW)~\cite{Greenwood1966RSPA} introduced a statistical
asperity model, relating the real contact area to the distribution of summit
heights.
Persson~\cite{Persson2001JCP} took a totally different approach: reformulating
the rough contact problem as a diffusion process in contact pressure, which
yielded closed-form predictions for the real contact area~\cite{Persson2002EPJE},
the interfacial gap~\cite{Persson2007PRL}, and the elastic
stiffness~\cite{Pastewka2013PRE} that can be linked to the height power spectrum.
The theory has been extensively validated against both finite-element
simulations~\cite{Hyun2004PRE} and experiments~\cite{Lorenz2008JPCM}, and has
since been extended to adhesion~\cite{Persson2002EPJE},
viscoelasticity~\cite{Persson2004JCP}, electroadhesion and
friction~\cite{Persson2018JCP,Persson2021JPCM}, and fluid leakage at
interfaces~\cite{Persson2022TL,Xu2026TI}.

On the other hand, M\"{u}ser established a rigorous field-theoretical framework
for rough elastic contacts~\cite{Mueser2008PRL}.
In this approach, the pressure distribution is obtained through a statistical
cumulant expansion,  which recasts Persson's theory as the leading-order and
provides a systematic route to higher-order corrections.
Beyond contact forces, the same field-theoretical framework also characterizes
the interfacial gap.
Zhou et al.~\cite{Zhou2026TI-2} extended the cumulant expansion to the gap
field, deriving an explicit analytical relation between the mean gap and applied
pressure together with a convection-diffusion equation for the scale-dependent
gap distribution.

Translating these theoretical insights into quantitative predictions requires
numerical methods capable of resolving the elastic field at the contacting
interface with sufficient resolution.
The finite element method (FEM)~\cite{Wriggers2003CM} accommodates
finite geometries and nonlinear constitutive laws but incurs prohibitive
volumetric meshing costs for problems involving large substrates and fine
surface discretizations.
Boundary element methods (BEM)~\cite{Putignano2014PM} reduce dimensionality by
one through surface Green's functions~\cite{Menga2014RSPA,Monti2021PRE}, and the
Green's function molecular dynamics (GFMD) method further accelerates the
computation to $\mathcal{O}(N\log N)$ scaling per iteration through fast
Fourier transform, making it the method of choice for large-scale rough contact
simulations on elastic slab~\cite{Campana2006PRB,Zhou2019PRB}.

In recent years, machine learning has introduced new methods for contact
mechanics problems.
Specifically, artificial neural networks (ANNs)~\cite{Kalliorinne2021FME,
Suman2025JT} have been used to create data-driven mappings from surface
topography to contact parameters, offering advantages in predicting real contact
area and its relationship with load.
However, purely data-driven models can violate fundamental mechanical laws due
to a lack of physical consistency.

This has motivated the development of algorithms that combine physical
knowledge with machine learning.
For example, Maghami et al.~\cite{Maghami2025IJSS} combined analytical
predictions with data-driven models trained on boundary-element simulations
to predict pull-off forces and work to pull-off in viscoelastic spherical
contacts.
Physics-informed neural networks (PINNs)~\cite{Schmidt2009S,Bongard2007PNAS,
Brunton2016PNAS} address this by embedding physical principles, typically
governing equations, into the loss function during
training~\cite{Raissi2019JCP,Chen2021NC}.
Zhou and Song demonstrated that PINNs can predict contact stress distributions
and relative contact areas for rough surfaces by using Persson’s diffusion
equation as a physics constraint, achieving accuracy within $0.5\%$ of GFMD
benchmarks, even when extrapolating beyond training data~\cite{Zhou2025TL}.
Later work extended this method to predict gap distributions under
partial-contact conditions where analytical solutions are not
available~\cite{Zhou2026TI-1}.

Furthermore, it has been demonstrated that the physical constraint implemented
in loss function does not necessarily be a differential equation.
Alternatively, it can also take the form of direct energy
minimization~\cite{Li2026JMPS}.
Bai et al. provides a compelling example in this direction, addressing the large
deformation and material nonlinearity in finite-body contact problems involving
smooth surfaces~\cite{Bai2025CMAME}.

Despite this, energy-based PINNs often exhibit lower computational efficiency
than traditional BEM when applied to contact mechanics.
This inefficiency can be attribute to two factors.
First, the composite energy-based loss, which sums the total potential energy
and boundary-condition penalty terms, gives rise to stiff gradient-flow
dynamics, producing strongly unbalanced back-propagation gradients across
different loss components~\cite{Wang2021JSC,He2026ARXIV}.
Second, and more fundamentally, a spectral-level issue underlies this gradient
imbalance.
It has been mentioned that \cite{Rahaman2019PMLR,Tancik2020ANIPS} standard neural
networks exhibit an inherent low-frequency bias, namely, the network parameterization
itself causes high-frequency Fourier components to be learned far more slowly
than low-frequency ones.

In contact mechanics, this difficulty is further intensified by the
Fourier-space structure of the elastic stiffness operator, which scales with the
wave number and thus intrinsically amplifies short-wavelength modes while
suppressing long-wavelength ones.
Short-range adhesive interactions introduce additional sharply localized
features, injecting yet higher-frequency content into the solution that further
widens the gap between fast and slow convergence modes~\cite{Zhou2019PRB}.

In this work, we would like to address the spectral stiffness challenge by
introducing a mass weighting (MW) spectral preconditioner designed specifically
for energy-minimizing PINN contact simulations.
The preconditioner operates in Fourier space prior to gradient back-propagation:
loss gradients are transformed to the Fourier domain, reweighted by a
carefully designed spectral mass weighting function that amplifies
low-wavenumber contributions while attenuating sub-grid-scale noise through an
embedded spectral low-pass filter, and then inverse-transformed back to real
space before being passed to the optimizer.

In the following, we will demonstrate the proposed mass weighting preconditioner
on one-dimensional adhesive line-contact problems with a cylindrical indenter
and randomly rough surfaces with Morse potential, using up to $n_x=2048$ spatial
discretization points, and compare its performance against unconditioned training.
We expect that, through comparisons against unconditioned training, the
mass weighting optimization will significantly accelerate convergence of the
macroscopic deformation field, yield smooth and oscillation-free contact-stress
profiles on fine grids, and maintain robust training dynamics without manual
tuning of the preconditioner parameters.

The remainder of this paper is organized as follows:
The model and methods investigated in this study are introduced in
Section~\ref{sec:model-method}.
The related results and discussions are presented in Section~\ref{sec:result}
and conclusions would be sketched in Section~\ref{sec:conclusion}.

\section{Model and method}\label{sec:model-method}

\subsection{Model}\label{sec:model}

\begin{figure*}[h!]
\centering
\includegraphics[width=0.95\linewidth]{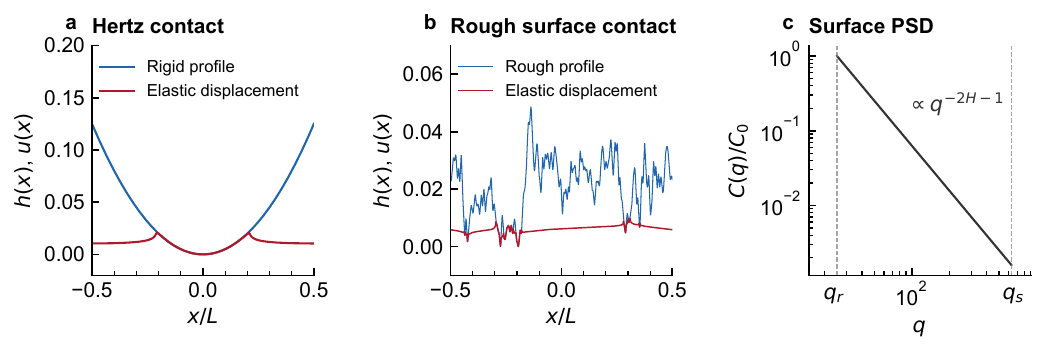}
\caption{%
\textbf{(a)} Schematic of Hertz contact: a rigid parabolic indenter (blue) and the
corresponding elastic half-space (red) under a nominal pressure.
\textbf{(b)} Schematic of randomly rough surface contact: the fractal surface profile
(blue) and its elastic deformation (red) under a nominal pressure.
\textbf{(c)} Power spectral density (PSD) of the rough surface,
$C(q) \propto q^{-2H-1}$ with $H=0.5$, shown on log-log scales over
the wavenumber range $[q_{\mathrm r}, q_{\mathrm s}]$.
}
\label{fig:model}
\end{figure*}

We consider an adhesive line-contact problem in the plane-strain regime,
as illustrated schematically in Fig.~\ref{fig:model}.
The rigid counter-surface is either a smooth parabolic profile of radius
$R_{\mathrm c}$, representing the classical Hertz indenter, whose profile is
\begin{equation}\label{eq:hertz_profile}
    h(x) = \frac{x^2}{2R_{\mathrm c}},
\end{equation}
or a randomly rough surface whose statistical topography is prescribed by the
height power spectral density (PSD)
\begin{equation}\label{eq:psd}
    C(q) = C_0\left( \frac{q}{q_{\mathrm r}} \right)^{-2H-1},
\end{equation}
where $H$ is the Hurst exponent, $q$ the wave number restricted to the interval
$[q_{\mathrm r}, q_{\mathrm s}]$ with $q_{\mathrm r} = 2\pi/\lambda_{\mathrm r}$
the roll-off and $q_{\mathrm s} = 2\pi/\lambda_{\mathrm s}$ the short-wavelength
cutoff, and $C_0$ a prefactor fixed by the root-mean-square height or gradient.
In this study, we would like to fix the root-mean-square gradient to unity by
default.
The Fourier coefficients of the rough surface height are constructed from
$\tilde h(q) = \sqrt{C(q)/L}\,\exp(i 2\pi X_q)$ with $X_q$ a random variable
uniform in $[0,1]$, and the real-space profile $h(x)$ is obtained by inverse
Fourier transform, which reads
\begin{equation}
    h(r) = \frac{L}{2\pi} \int {\mathrm d}q~\tilde h(q) \exp(iqx)
\end{equation}

Both surfaces are fixed in space and well designed so that the minimum of $h(x)$
is zero.
An elastic half-space is pressed against a rigid counterface under a nominal external
pressure $p_0$ applied in the center-of-mass mode.
The half-space is characterized by an effective contact
modulus $E^* = E / (1-\nu^2)$, where $E$ represents the elastic modulus
and $\nu$ the Poisson ratio.
The adhesive interaction between the two surfaces is described by a Morse
potential.
The surface of the half-space undergoes a normal displacement field $u(x)$,
defined on a one-dimensional domain $x \in [-L/2, L/2]$ discretized into $n_x$
uniformly spaced grid points.
If not mentioned explicitly, the system size $L$, effective modulus
$E^*$ are fixed to unity by default.

The total potential energy $\Pi$ of the system comprises three contributions:
\begin{equation}\label{eq:total_energy}
    \Pi = U_{\mathrm{el}} + U_{\mathrm{ext}} + U_{\mathrm{int}},
\end{equation}
where $U_{\mathrm{el}}$ is the elastic strain energy stored in the half-space,
$U_{\mathrm{ext}}$ the work done by the external pressure, and $U_{\mathrm{int}}$
the interaction energy across the interface.
For a semi-infinite elastic solid, the elastic energy can be compactly
expressed in Fourier space through the spectral stiffness
operator~\cite{Persson2001JCP,Prodanov2013TL}:
\begin{equation}
\label{eq:elastic_energy}
U_{\mathrm{el}} = \frac{L}{2} \sum_{q} \frac{qE^*}{2} \vert \tilde u(q) \vert^2,
\end{equation}
where $\tilde{u}(q)$ denotes the Fourier transform of $u(x)$.
The factor $qE^*/2$ is the Fourier representation of the elastic Green's
function for a half-space and encodes the scale-dependent stiffness of the
substrate, that is, short-wavelength deformations (large $q$) incur higher
elastic energy cost than long-wavelength ones (small $q$).
The real-space elastic stress at the surface, $\sigma(x)$, is recovered from the
inverse Fourier transform of the spectral stress
$\tilde{\sigma}(q) = -(qE^*/2) \tilde{u}(q)$.

The work performed by the nominal pressure $p_0$ acting over the entire nominal
contact area is
\begin{equation}\label{eq:external_work}
    U_{\mathrm{ext}} = -p_0 L \tilde{u}(0),
\end{equation}
where $\tilde u(0)$ denotes the center-of-mass mode of the elastic displacement.
The interaction forces between the two surfaces are modeled by a Morse potential,
which captures both short-range repulsion and long-range
attraction~\cite{Zhou2019PRB}:
\begin{equation}\label{eq:morse}
    \gamma(g) = \gamma_0 \left[ \exp\left(-\frac{2g}{\rho}\right)
                     - 2\exp\left(-\frac{g}{\rho}\right) \right],
\end{equation}
where $\gamma_0$ is the work of adhesion, $\rho$ is the characteristic range of
the interaction and $g$ denotes the interfacial gap which is determined by
\begin{equation}
    g(x) = h(x) - u(x).
\end{equation}
The total interfacial energy is obtained by integrating the energy density over the domain:
\begin{equation}\label{eq:interaction_energy}
    U_{\mathrm{int}} = L \sum_i \gamma\big[g(x_i)\big].
\end{equation}

The finite-range interaction in Eq.~\ref{eq:morse} retains adhesive
forces at nonzero gaps; the energy-based formulation therefore does
not itself impose the zero-range JKR approximation.
For comparison with GFMD, we use the same operational contact
criterion, $g(x_i)\leq2\rho$, and calculate the relative contact area as
\begin{equation}\label{eq:relative_contact_area}
    \frac{A_{\mathrm c}}{A_0}
    = \frac{1}{n_x}\sum_{i=1}^{n_x}
      \mathbf{1}\!\left[g(x_i)\leq2\rho\right],
\end{equation}
where $\mathbf{1}$ is the indicator function.
For a single contiguous line-contact region, the corresponding
contact half-width is $c=(L/2)(A_{\mathrm c}/A_0)$.
This gap-based measure includes a finite positive-gap region and
should be distinguished from the repulsive contact area.

\subsection{Method}\label{sec:method}

\begin{figure*}[h!]
\centering
\includegraphics[width=0.95\linewidth]{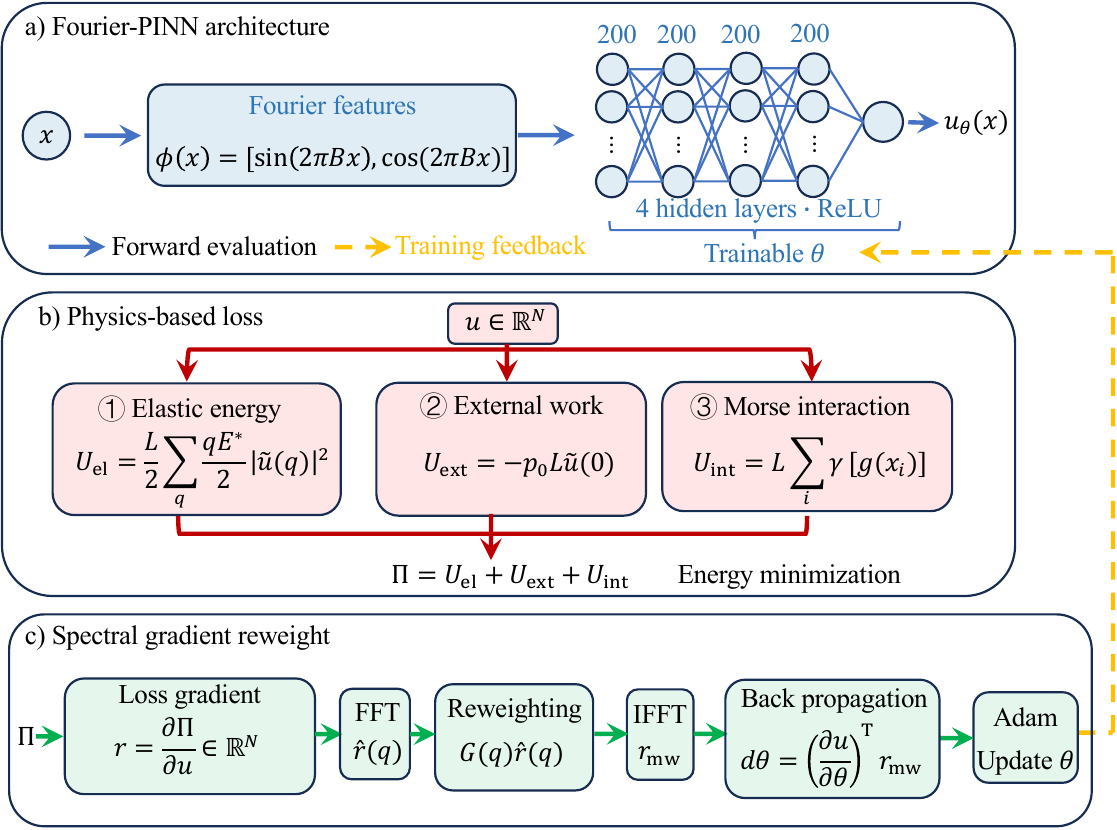}
\caption{%
\textbf{(a)} Spatial coordinates are mapped through fixed Fourier features and four fully
connected hidden layers to predict displacements.
\textbf{(b)} The displacement field and prescribed physical inputs determine the elastic,
external and interaction energies, whose sum defines the loss.
\textbf{(c)} Loss gradients undergo spectral mass weighting and low-pass filter before
back-propagation and Adam parameter updates.
}
\label{fig:fourier_pinn}
\end{figure*}

The learning task is to find the displacement field $u(x)$ that minimizes the
total potential energy $\Pi[u]$ defined in Eq.~\ref{eq:total_energy}, given
the surface profile $h(x)$, the nominal pressure $p_0$, and the material
parameters $(E^*, \gamma_0, \rho)$.
Formally, we seek the optimal network parameters $\theta^*$ such that
\begin{equation}\label{eq:optimization}
    \theta^* = \arg\min_\theta \; \Pi[u_\theta].
\end{equation}
Unlike conventional PINNs that balance PDE residuals with boundary-condition
penalties, this energy-minimizing formulation requires only a single scalar loss
function and does not require explicit enforcement of boundary conditions.

No precomputed simulations are used as training data.
For each prescribed surface profile, loading condition, and set of material parameters, the
network is optimized directly by minimizing the total potential energy evaluated on the spatial
grid.
No reference displacement or stress labels are required.

The spatial coordinate $x$ is encoded through a Fourier feature mapping, then
passed through a feed-forward neural network to produce the displacement
$u_\theta(x)$.
The total potential energy $\Pi$ is assembled by evaluating the elastic energy
in Fourier space (Eq.~\ref{eq:elastic_energy}), the external work
(Eq.~\ref{eq:external_work}), and the Morse potential
(Eq.~\ref{eq:interaction_energy}) in real space.
Before back-propagation, the gradient $\partial\Pi / \partial u$ is transformed
to Fourier space, reweighted by a mass-weighting factor and a spectral low-pass
filter, and transformed back to real space.
The preconditioned gradient then flows through the network via standard
automatic differentiation, and the Adam optimizer updates $\theta$.

The displacement field $u(x)$ is parameterized by a fully connected
feed-forward neural network $u_\theta(x)$ with trainable parameters $\theta$.
To resolve the sharp displacement gradients that arise near contact edges, the
input coordinate $x$ is first mapped to a set of Fourier
features~\cite{Tancik2020ANIPS}:
\begin{equation}\label{eq:fourier_features}
    \bm{\phi}(x) = \big[ \sin(2\pi \mathbf{B} x),\; \cos(2\pi \mathbf{B} x) \big],
\end{equation}
where $\mathbf{B}$ is a random Gaussian matrix with entries drawn
from $\mathcal{N}(0, s^2)$ and $s$ is a scale factor set to the
maximum wave number resolved by the grid, $s \approx \pi L/ n_x$.
The Fourier feature matrix $\mathbf{B}$ remains fixed during training.
The feature vector $\bm{\phi}(x)$ is passed through four fully connected hidden layers ($n_l=4$),
each containing 200 neurons ($n_h=200$) with ReLU activation, followed by a linear output layer
with one neuron predicting the displacement.
Only the weights and biases of the fully connected layers are trainable.
The number of Fourier features is scaled linearly with the grid size to maintain sufficient
spectral resolution.

A direct gradient descent on $\Pi$ with respect to $\theta$ suffers from a
spectral stiffness imbalance, that is, short-wavelength displacement modes couple to the
elastic energy through the factor $qE^*/2$, which grows linearly with wave
number, whereas the adhesive interaction stiffness is wavenumber-independent.
Consequently, high-$q$ modes dominate the gradient updates and stall convergence
of the macroscopic deformation.
To solve this problem, we apply a spectral preconditioner to the displacement
gradient $\partial\Pi / \partial u$ before back-propagation.
Let $\tilde{r}(q) = \mathcal{F}[\partial\Pi / \partial u]$ denote the Fourier
transform of the real-space gradient.
Each Fourier mode is reweighted by a mass-weighting factor~\cite{Zhou2019PRB}:
\begin{equation}\label{eq:mass_weight}
    w(q) = \left(
        \frac{E^* q_{\max} / 2}{\sqrt{k_0^2 + (qE^*/2)^2}}
    \right)^{\alpha},
\end{equation}
where $k_0 = \xi (E^* q_{\max} / 2)$ is a low-$q$ stiffness floor with
$\xi \ll 1$, and the exponent $\alpha \in (0,1]$ controls the weighting strength,
which is fixed to $0.5$ in this study.
The reweighted factor $w(q)$ is clamped to $[w_{\min}, w_{\max}]$ to prevent
extreme gain values.
Additionally, a built-in spectral low-pass filter suppresses sub-grid noise:
\begin{equation}\label{eq:lowpass}
    f(q) = \frac{1}{1 + (q / q_{\mathrm{c}})^{\beta}},
\end{equation}
where $q_{\mathrm{c}}$ is set proportional to the reciprocal of the interaction
range and $\beta$ controls the filter roll-off steepness.
The combined preconditioning weight is $G(q) = w(q) \, f(q)$, and the
preconditioned real-space gradient is obtained by inverse transform:
\begin{equation}\label{eq:precond_grad}
\frac{\delta\Pi}{\delta u} \bigg\vert_{\mathrm{precond}}
= \mathcal{F}^{-1}\big[ G(q) \, \tilde{r}(q) \big].
\end{equation}
Back-propagation through the network then uses this preconditioned gradient,
effectively rebalancing the spectral contributions so that long-wavelength modes
receive amplified gradient signals.

The network is trained using the Adam optimizer with an initial learning rate of
$10^{-4}$.
The displacement field is initialized to the uniform value $u(x) = -\rho\ln 2$.
For large grids, the displacement is evaluated in mini-batches to reduce memory
overhead, while the Fourier-space operations are performed on the full assembled
field to preserve spectral accuracy.
To ensure training stability, the preconditioner is activated only after a
warm-up phase of $N_{\mathrm{warm}} = 100 \sim 200$ plain-gradient steps.
For the Hertzian contact benchmark with 2048 grid points, a run of
500 Adam iterations took 17.8~s on a compute node equipped with an
NVIDIA RTX 4090D GPU with 24~GB of memory.
Mass weighting was activated after 200 warm-up iterations.
After training, the median time to predict the displacement field
over all 2048 grid points was 0.723~ms, measured over 100 evaluations
after 20 warm-up evaluations.
GPU execution was synchronized at the timing boundaries.
The inference measurement excludes FFT-based stress reconstruction,
data transfer to the CPU, and file output.
This timing refers to evaluation of the trained solution for the
prescribed contact problem; changing the physical conditions
generally requires further optimization.

At convergence, the contact stress distribution is obtained from the spectral
stress--displacement relation:
\begin{equation}\label{eq:stress}
    \sigma(x) = \mathcal{F}^{-1}\big[ -(E^*/2) q \, \tilde{u}_\theta(q) \big],
\end{equation}
with the zero-wavenumber component corrected by the nominal pressure,
$\tilde{\sigma}(0) \leftarrow \tilde{\sigma}(0) - p_0$.
Computing the stress in Fourier space avoids the grid-sensitivity inherent in
real-space numerical differentiation of $u_\theta(x)$ and is consistent with the
spectral form of the elastic energy.

In summary, the present formulation is restricted to one-dimensional
line-contact problems under plane-strain conditions with a uniform grid
spacing $\Delta x$.
The elastic half-space is assumed semi-infinite and homogeneous, and the
adhesive interaction is modeled by the Morse potential with a range
$\rho \ll L$.
The spectral preconditioner assumes that the gradient spectrum is dominated by
the elastic stiffness contribution; its effectiveness for other interaction
potentials or three-dimensional geometries remains to be investigated.

\section{Results}\label{sec:result}

\begin{figure}[h!]
\centering
\includegraphics[width=0.95\linewidth]{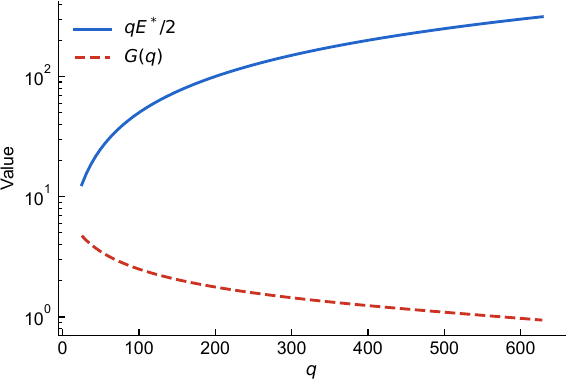}
\caption{%
The elastic stiffness $qE^*/2$ (blue) increases linearly with wavenumber,
spanning more than one order of magnitude across the spectral band.
The mass-weighting function $G(q)$ (red dashed) compensates this imbalance
by amplifying low-$q$ gradients and attenuating high-$q$ contributions.
}
\label{fig:elastic_kernel}
\end{figure}

The elastic energy expression reveals the central challenge for gradient-based
training: in Fourier space the stiffness kernel $E(q) = qE^*/2$ grows linearly
with wave number $q$ (Eq.~\ref{eq:elastic_energy}).
Figure~\ref{fig:elastic_kernel} shows this kernel spanning across the spectral
bandwidth $[q_{\mathrm r}, q_{\mathrm s}]$ of a multiscale rough surface.
This spectral heterogeneity means that high-$q$ displacement modes experience
elastic gradients that dominate those of long-wavelength modes by the same
factor, causing Adam updates to be driven primarily by short-wavelength
fluctuations while the macroscopic deformation geometry remains under-optimized.
The mass-weighting function $G(q)$ designed to compensate this imbalance, is
depicted in Fig.~\ref{fig:elastic_kernel} (red dashed): at low $q$ it applies a
gain of approximately 4 to 5, amplifying the signal for macroscopic modes, while
at high $q$ it decays to suppress spurious short-wavelength dynamics.

\begin{figure}[h!]
\centering
\includegraphics[width=0.95\linewidth]{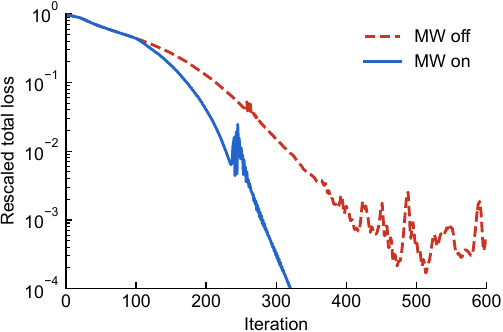}
\caption{%
Resccaled loss (log scale) versus Adam iteration for the Hertzian contact
problem.
Mass weighting off (red dashed): the loss stalls at $\sim 5 \times 10^{-4}$.
Mass weighting on (blue solid): the loss converges to machine zero within
400 iterations.
The preconditioner is activated at iteration~200 (warm-up).
}
\label{fig:rescaled_loss}
\end{figure}

To quantify the effect of spectral preconditioning on convergence, we trained
the Fourier PINN for 600 Adam iterations on a Hertzian contact problem
with $R_{\mathrm c} = 1.0$, $p_0 / E^* = -1.0 \times 10^{-3}$,
$\gamma_0 = 1.0\times 10^{-3}$ and $\rho / L = 1.58 \times 10^{-3}$ with
mass weighting enabled and disabled.
Figure~\ref{fig:rescaled_loss} reports the rescaled total
loss $(\mathcal{L} - \mathcal{L}_n)/(\mathcal{L}_0- \mathcal{L}_n)$,
where $\mathcal{L}_0$ is the initial loss value, and $\mathcal{L}_n$
the final loss value, on a logarithmic scale.
Without preconditioning, the loss decreased to approximately $5 \times 10^{-4}$
of its initial value and then stalled; further iterations produced no meaningful
improvement (red dashed).
With mass weighting activated after a 200-step warm-up phase, the loss decayed
monotonically and reached machine-zero residual within 400 iterations (blue
solid).
The visible kink in the convergence curve at iteration~200 coincides with
activation of the preconditioner, confirming that the acceleration is directly
attributable to the spectral rebalancing.

\begin{figure}[h!]
\centering
\includegraphics[width=0.95\linewidth]{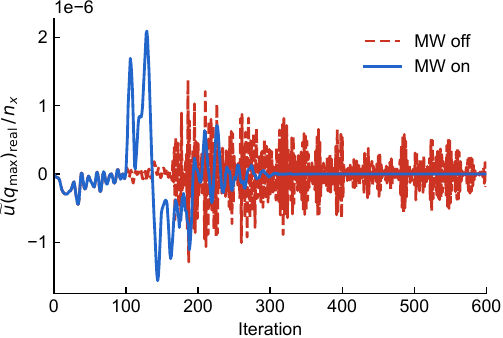}
\caption{%
Amplitude of the highest-wavenumber Fourier component,
$\widetilde{u}(q_{\max})_{\mathrm{real}} / n_x$, versus Adam iteration.
Mass weighting off (red dashed): persistent large-amplitude oscillations.
Mass weighting on (blue solid): rapid decay to zero after warm-up.
}
\label{fig:fastest_mode}
\end{figure}

To understand why the unpreconditioned training stalls, we monitored the
amplitude of the fastest Fourier mode,
$\tilde{u}(q_{\max})_{\mathrm{real}} / n_x$, which is the mode most strongly
affected by spectral stiffness imbalance.
Figure~\ref{fig:fastest_mode} shows this mode's evolution throughout training.
Without preconditioning, the fastest mode oscillated with persistent large
amplitude and never settled to a steady value (red dashed), indicating that
high-$q$ gradient noise continuously perturbed the displacement field and
prevented the network from converging to the energy minimum.
With preconditioning, the same mode amplitude decayed rapidly to zero within the
warm-up phase and remained stable thereafter (blue solid), confirming that the
mass weighting effectively suppresses the spurious high-frequency dynamics that
would otherwise dominate the gradient signal.

\begin{figure}[h!]
\centering
\includegraphics[width=0.95\linewidth]{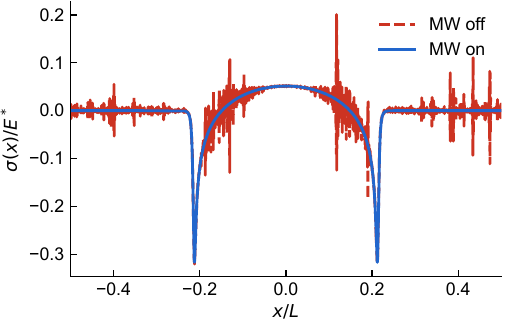}
\caption{%
Spatial stress distribution $\sigma(x)/E^*$ for the Hertzian contact problem.
Mass weighting off (red dashed): noisy, unphysical stress with spurious
oscillations.
Mass weighting on (blue solid): smooth, physically correct stress with
resolved compressive contact patch.
}
\label{fig:stress_profile}
\end{figure}

The physical consequence of stable spectral convergence is directly visible
in the spatial stress field.
Figure~\ref{fig:stress_profile} compares the contact stress
$\sigma(x)/E^*$ obtained from the PINN trained with and without mass weighting.
The unpreconditioned solution exhibits a noisy, unphysical stress profile with
spurious high-frequency oscillations spanning the entire domain (red dashed).
In contrast, the preconditioned solution yields a smooth stress field that
correctly resolves compressive contact patch separated by stress-free
non-contact regions (blue solid).
The difference is not quantitative but qualitative: without spectral
preconditioning, the PINN fails to learn the physically correct contact solution
even though the energy functional is identical in both cases.

Having established that spectral preconditioning is essential for stable
training, we validate the accuracy of the converged PINN solution against the
Green's function molecular dynamics (GFMD) reference for the classical Hertz
contact problem, that is, a smooth parabolic indenter on an elastic
half-space (Fig.~\ref{fig:model}a).
Three far-field pressure levels were tested:
$p_0/E^* = -0.01$ (tension),
$p_0/E^* = 0$  (zero external load), and
$p_0/E^* = 0.01$ (compression).

\begin{figure}[h!]
\centering
\includegraphics[width=0.95\linewidth]{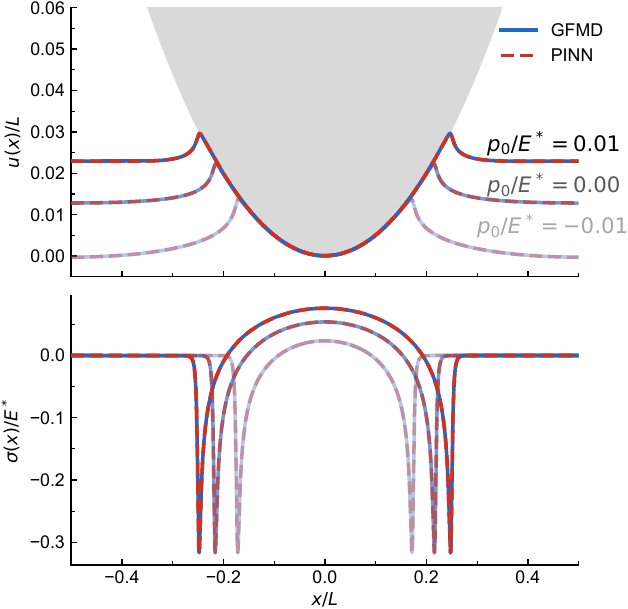}
\caption{%
Top panel: elastic displacement $u(x)/L$.
Bottom panel: contact stress $\sigma(x)/E^*$.
Three far-field pressures $p_0/E^* \in \{-0.01,\,0,\,0.01\}$ are shown with
increasing opacity.
GFMD reference (blue solid); PINN prediction (red dashed).
}
\label{fig:hertz}
\end{figure}

Figure~\ref{fig:hertz} shows the elastic displacement $u(x)$ (top panel)
and contact stress $\sigma(x)/E^*$ (bottom panel) for all three cases.
Opacity encodes the pressure level, increasing from $p_0 = -0.01$ (faint) to
$p_0 = 0.01$ (opaque).
The PINN predictions (red dashed) are indistinguishable from the GFMD reference
(blue solid) across the entire spatial domain, for both displacement and stress,
at all three pressure levels.
This agreement demonstrates that the Fourier PINN with mass-weighted spectral
preconditioning correctly solves the linear elastic contact problem with an
imposed smooth rigid profile.

To complement the displacement and stress comparisons, we examine
the load dependence of the contact half-width for the smooth
parabolic indenter.
We introduce the characteristic scales
\begin{equation}\label{eq:adhesive_contact_scales}
    c^*=\left(\frac{R_{\mathrm c}^2\gamma_0}{E^*}\right)^{1/3},
    \qquad
    p^*=\frac{1}{L}
    \left(E^*R_{\mathrm c}\gamma_0^2\right)^{1/3},
\end{equation}
and define $\bar c=c/c^*$ and $\bar p=p_0/p^*$, with compression
taken as positive.
The JKR solution for line contact then reads
\begin{equation}\label{eq:jkr_line_contact}
    \bar p=\frac{\pi}{4}\bar c^2-\sqrt{2\pi\bar c}.
\end{equation}

\begin{figure}[h!]
\centering
\includegraphics[width=0.95\linewidth]{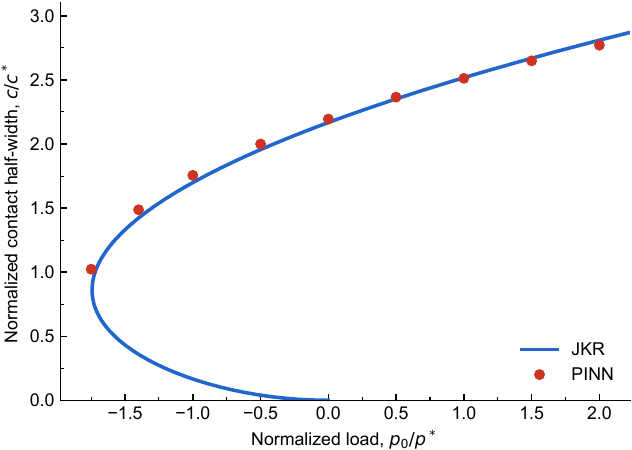}
\caption{%
Normalized contact half-width versus normalized load for adhesive
line contact at $\mu_\rho=6.33$.
Fourier-PINN predictions (red filled circles) are compared with the
JKR solution (blue solid line), including its unstable lower branch
under load control.
}
\label{fig:jkr_vs_pinn}
\end{figure}

The adhesion regime is characterized using the Morse interaction
length through the Tabor-type parameter
\begin{equation}\label{eq:tabor_morse}
    \mu_\rho=
    \left(\frac{R_{\mathrm c}\gamma_0^2}
    {E^{*2}\rho^3}\right)^{1/3}.
\end{equation}
For this benchmark, $R_{\mathrm c}=1$, $E^*=1$,
$\gamma_0=10^{-3}$, and $\rho=1.58\times10^{-3}$,
giving $\mu_\rho\simeq6.33$.
Figure~\ref{fig:jkr_vs_pinn} shows that the Fourier-PINN contact
half-widths closely follow the JKR reference over the investigated
load range, supporting the method in this short-range adhesion regime.
The unstable lower branch of the analytical curve is included for
reference and is not a loading or unloading trajectory under load control.

Comparison with the DMT limit requires additional care.
The criterion $g\leq2\rho$ includes part of the positive-gap region,
whose contribution to the measured contact size can become more
pronounced as $\rho$ increases.
Moreover, the Morse potential couples the attractive range to the
repulsive stiffness: its curvature at equilibrium is
$\gamma''(0)=2\gamma_0/\rho^2$.
Increasing $\rho$ at fixed $\gamma_0$ therefore reduces
$\mu_\rho$ while also softening the repulsive interaction, and does
not by itself ensure the hard-contact configuration underlying
the DMT approximation.
M{\"u}ser~\cite{Mueser2014BJN} treated non-penetration through a
hard-wall constraint separately from the finite-range interaction.
A hard-wall constraint, or sufficiently stiff short-range repulsion,
would provide a suitable basis for a dedicated DMT comparison,
together with consistent contact definitions and checks of
resolution, domain size, and convergence.
The present benchmark does not establish accuracy in the DMT limit.

For multiscale rough surfaces, the relevant local curvature depends
on the observation scale, so a single Tabor parameter cannot
characterize all roughness scales~\cite{Papangelo2026Elsevier}.
The value reported above therefore applies to the smooth benchmark.
The rough-contact comparisons below assess agreement with GFMD
for the prescribed surface spectra and interaction parameters.

We next apply the method to contact with a fractal rough surface whose PSD is
given by Eq.~\ref{eq:psd} with $H = 0.5$, containing roughness over two decades
of wavelength from $\lambda_{\mathrm r} = 0.25L$ to
$\lambda_{\mathrm s} = 0.01L$ (Fig.~1c).
Three far-field pressures were applied, spanning the transition from tension
($p_0/(E^*\bar{g}) = -1.5 \times 10^{-3}$ and $-1.0 \times 10^{-3}$) to
compression ($p_0/(E^*\bar{g}) = 1.0 \times 10^{-2}$), where
$\bar{g}$ denotes the root-mean-square gradient of height of rough surface.
These pressures were chosen to sample the regime in which the real contact area
is a small fraction of the nominal area, where multi-scale roughness effects are
most pronounced.

\begin{figure*}[h!]
\centering
\includegraphics[width=0.95\linewidth]{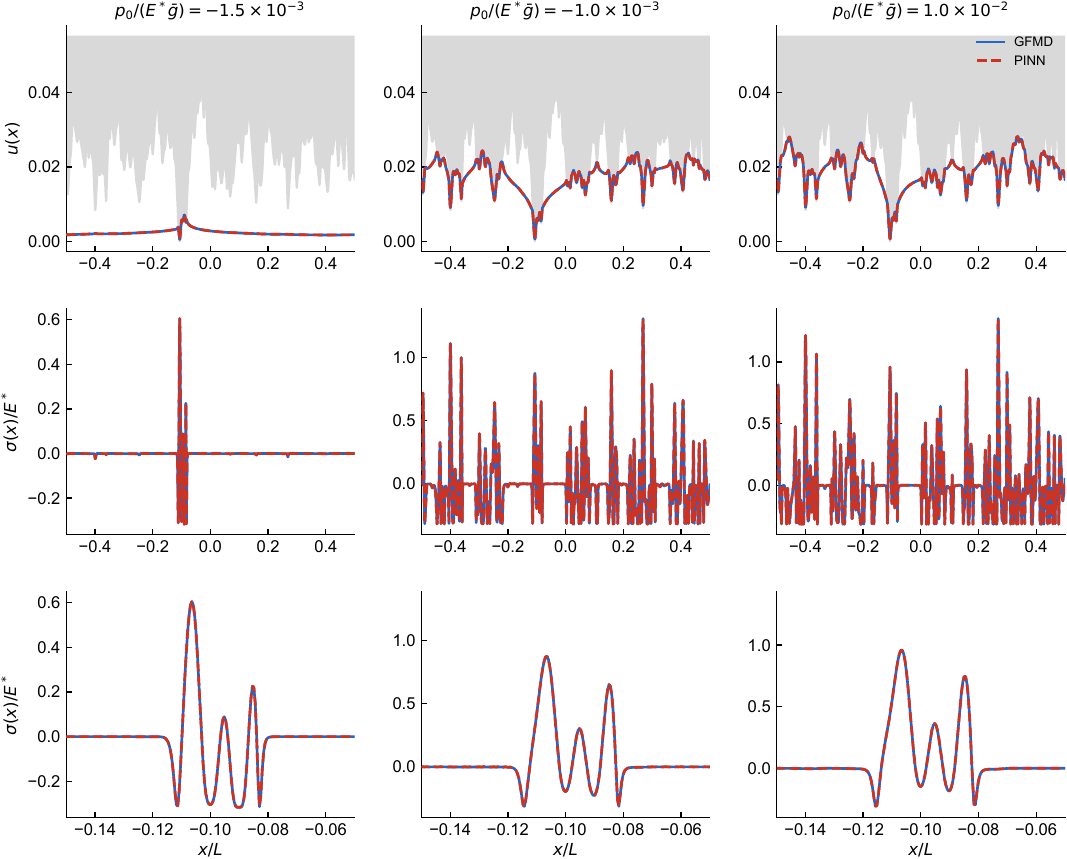}
\caption{%
Top row: elastic displacement $u(x)$.
Middle row: full-domain contact stress $\sigma(x)/E^*$.
Bottom row: zoomed stress over $x \in [-0.15, -0.05]$.
Columns correspond to three far-field pressures
$p_0/(E^*\bar{g}) \in \{-1.5\times10^{-3},\,-1.0\times10^{-3},\,1.0\times10^{-2}\}$.
GFMD reference (blue solid); PINN prediction (red dashed).
}
\label{fig:fractal}
\end{figure*}

Figure~\ref{fig:fractal} presents the elastic displacement $u(x)$ (top row),
full-domain contact stress $\sigma(x)/E^*$ (middle row), and a zoomed-in stress
profile over $x \in [-0.15, -0.05]$ (bottom row) for all three pressures.
Across the full pressure range, the PINN predictions (red dashed) closely track
the GFMD reference (blue solid) in both the shape and amplitude of the elastic
displacement field.
The contact stress captures the defining feature of rough-surface contact:
isolated compressive peaks separated by stress-free gaps at the two lower
pressures, reflecting the sparse, asperity-level contact morphology, and a
transition to a nearly continuous compressive stress distribution at the highest
pressure.

The zoomed view confirms that the PINN resolves individual contact patches with
quantitative accuracy, correctly reproducing both the positions and amplitudes
of stress peaks as predicted by GFMD.
These results establish that the Fourier PINN with spectral preconditioning
solves rough-surface contact mechanics with fidelity sufficient to capture the
multi-scale stress structure that underpins tribological properties such as
real contact area, friction, and interfacial stiffness.

To provide a quantitative complement to the visual comparisons in
Figs.~\ref{fig:hertz} and~\ref{fig:fractal}, we define the root-mean-square
error (RMSE) and the normalized RMSE (NRMSE) for a field $f \in \{u, \sigma\}$ as
\begin{equation}\label{eq:rmse}
    \mathrm{RMSE}_f = \sqrt{\frac{1}{n_x}\sum_{i=1}^{n_x}
        \big[ f_{\mathrm{PINN}}(x_i) - f_{\mathrm{GFMD}}(x_i) \big]^2},
\end{equation}
\begin{equation}\label{eq:nrmse}
    \mathrm{NRMSE}_f = \frac{\mathrm{RMSE}_f}
        {\max_i f_{\mathrm{GFMD}}(x_i) - \min_i f_{\mathrm{GFMD}}(x_i)},
\end{equation}
where $n_x = 2048$ is the number of spatial grid points and $f_{\mathrm{GFMD}}$
denotes the GFMD reference solution.
Table~\ref{tab:rmse} reports these metrics for displacement and stress across
all six test cases.
For the Hertz problem, the NRMSE of displacement lies between
$2.1 \times 10^{-6}$ and $1.2 \times 10^{-4}$, while the stress NRMSE
ranges from $1.9 \times 10^{-5}$ to $4.1 \times 10^{-4}$.
For the fractal rough surface, the displacement NRMSE remains below
$2.6 \times 10^{-4}$ and the stress NRMSE below $1.1 \times 10^{-3}$ across
all pressures.
Across all six cases, the mean NRMSE is $6.5 \times 10^{-5}$ for displacement
and $2.5 \times 10^{-4}$ for stress, confirming that the PINN predictions are
quantitatively consistent with the GFMD reference to within a small fraction of
one percent of the field range.
The largest stress NRMSE ($1.0 \times 10^{-3}$, fractal case at
$p_0/(E^*\bar{g}) = -1.0 \times 10^{-3}$) coincides with the lowest applied
pressure, where contact patches are sparse and the stress field is dominated by
a few narrow peaks whose precise amplitude is most sensitive to the
discretization of the Fourier representation.
\begin{table*}[h!]
\centering
\caption{%
RMSE and NRMSE (RMSE normalized by the range of the GFMD field) for
displacement $u$ and contact stress $\sigma$.
}
\label{tab:rmse}
\small
\begin{tabular}{lcccc}
\hline
 & \textbf{RMSE}$_u$ & \textbf{NRMSE}$_u$ & \textbf{RMSE}$_\sigma$ & \textbf{NRMSE}$_\sigma$ \\
\hline
\multicolumn{5}{c}{\textit{Hertz contact}} \\
\ \ $p_0/E^* = -0.01$ & $1.68 \times 10^{-6}$  & $1.20 \times 10^{-4}$  & $1.41 \times 10^{-4}$  & $4.13 \times 10^{-4}$ \\
\ \ $p_0/E^* = 0$     & $4.77 \times 10^{-8}$  & $2.15 \times 10^{-6}$  & $6.97 \times 10^{-6}$  & $1.88 \times 10^{-5}$ \\
\ \ $p_0/E^* = 0.01$  & $5.56 \times 10^{-8}$  & $1.88 \times 10^{-6}$  & $1.71 \times 10^{-5}$  & $4.35 \times 10^{-5}$ \\[4pt]
\multicolumn{5}{c}{\textit{Fractal rough surface}} \\
\ \ $p_0/(E^*\bar{g}) = -1.5 \times 10^{-3}$ & $3.66 \times 10^{-9}$  & $5.56 \times 10^{-7}$  & $2.09 \times 10^{-6}$  & $2.26 \times 10^{-6}$ \\
\ \ $p_0/(E^*\bar{g}) = -1.0 \times 10^{-3}$ & $6.17 \times 10^{-6}$  & $2.60 \times 10^{-4}$  & $1.66 \times 10^{-3}$  & $1.02 \times 10^{-3}$ \\
\ \ $p_0/(E^*\bar{g}) = 1.0 \times 10^{-2}$  & $3.16 \times 10^{-8}$  & $1.15 \times 10^{-6}$  & $1.70 \times 10^{-5}$  & $1.02 \times 10^{-5}$ \\
\hline
\end{tabular}
\end{table*}

\section{Conclusions}\label{sec:conclusion}

We have presented a Fourier PINN framework for one-dimensional adhesive contact
mechanics that uses an energy-minimizing formulation and spectral mass-weighting
preconditioning to overcome the stiffness imbalance inherent in elastic
half-space problems.
The key finding is that spectral preconditioning is not merely helpful but
\emph{necessary} for stable convergence: without it, training stalls at a finite
residual loss and produces unphysical stress fields dominated by high-frequency
noise; with it, the loss reaches machine zero within a few hundred Adam
iterations and the solution converges to the GFMD reference with quantitative
accuracy.
Monitoring the fastest Fourier mode amplitude during training directly confirms
that the preconditioner acts by suppressing the spurious high-$q$ dynamics that
otherwise overwhelm the gradient signal.

The method was validated on two complementary contact problems.
For the Hertz indenter, the Fourier PINN reproduces the GFMD displacement and
stress fields across three far-field pressures spanning tension to
compression (Fig.~\ref{fig:hertz}).
For fractal rough surfaces with roughness defined by specified PSD,
the PINN captures the multi-scale stress structure, isolated compressive peaks
at low pressure, transitioning to a continuous stress distribution at high
pressure, in quantitative agreement with GFMD (Fig.~\ref{fig:fractal}).
In both cases the real-space predictions are indistinguishable from the
reference solution, establishing that energy-minimizing PINNs with spectral
preconditioning can solve rough-surface contact problems without requiring the
explicit Green's function integration or all-to-all coupling that conventional
boundary-element methods rely on.

The present work is limited to one-dimensional line-contact geometries, the
moderate Hurst exponent $H = 0.5$, and adhesive interactions modeled by the
Morse potential.
Extension to two-dimensional rough surfaces is the natural next step: the
Fourier-space formulation of the elastic energy generalizes directly to 2D,
while the spectral preconditioner depends only on the wave-number $q$ and
should transfer without structural modification.
The interaction-potential independence of the preconditioner also suggests
applicability to a broader class of interface models, including
Lennard-Jones-based adhesion and rate-dependent friction laws, provided the
non-elastic energy contributions remain smooth in real space.

\bmhead{Acknowledgments}

YZ acknowledges the financial supports from the National Natural Science
Foundation of China (Grant No.12402116) and the Opening fund of State Key Laboratory of Nonlinear
Mechanics (Grant No. LNM202510).
HS acknowledges the financial support from the Strategic Priority Research Program
of the Chinese Academy of Science through grant No. XDB0620101.
YX acknowledges the financial support from the Natural Science Foundation of Anhui
Province (Grant No. 2508085ME101).

\section*{Declarations}
The author declares that he has no conflict interest.

\section*{Data Availability}
The Fourier based PINN source code and related data files can be
found \href{here}{https://github.com/yunongzhou/fourier_pinn}.

\section*{Declaration of generative AI and AI-assisted technologies in the manuscript preparation process}
During the preparation of this manuscript, the authors used ChatGPT and DeepSeek
to assist with language editing and clarity improvement.
All content was carefully reviewed and revised by the authors, who take full
responsibility for the accuracy, originality, and integrity of the published work.

\bibliography{pinn_adh}

@article{Bai2025CMAME,
  title = {Energy-based physics-informed neural network for frictionless contact problems under large deformation},
  volume = {437},
  ISSN = {0045-7825},
  url = {http://dx.doi.org/10.1016/j.cma.2025.117787},
  DOI = {10.1016/j.cma.2025.117787},
  journal = {Computer Methods in Applied Mechanics and Engineering},
  publisher = {Elsevier BV},
  author = {Bai,  Jinshuai and Lin,  Zhongya and Wang,  Yizheng and Wen,  Jiancong and Liu,  Yinghua and Rabczuk,  Timon and Gu,  YuanTong and Feng,  Xi-Qiao},
  year = {2025},
  month = mar,
  pages = {117787}
}

@article{Bongard2007PNAS,
  title = {Automated reverse engineering of nonlinear dynamical systems},
  volume = {104},
  ISSN = {1091-6490},
  url = {http://dx.doi.org/10.1073/pnas.0609476104},
  DOI = {10.1073/pnas.0609476104},
  number = {24},
  journal = {Proceedings of the National Academy of Sciences},
  publisher = {Proceedings of the National Academy of Sciences},
  author = {Bongard,  Josh and Lipson,  Hod},
  year = {2007},
  month = jun,
  pages = {9943–9948}
}

@article{Brunton2016PNAS,
  title = {Discovering governing equations from data by sparse identification of nonlinear dynamical systems},
  volume = {113},
  ISSN = {1091-6490},
  url = {http://dx.doi.org/10.1073/pnas.1517384113},
  DOI = {10.1073/pnas.1517384113},
  number = {15},
  journal = {Proceedings of the National Academy of Sciences},
  publisher = {Proceedings of the National Academy of Sciences},
  author = {Brunton,  Steven L. and Proctor,  Joshua L. and Kutz,  J. Nathan},
  year = {2016},
  month = mar,
  pages = {3932–3937}
}

@article{Campana2006PRB,
  title = {Practical Green’s function approach to the simulation of elastic semi-infinite solids},
  volume = {74},
  ISSN = {1550-235X},
  url = {http://dx.doi.org/10.1103/PhysRevB.74.075420},
  DOI = {10.1103/physrevb.74.075420},
  number = {7},
  journal = {Physical Review B},
  publisher = {American Physical Society (APS)},
  author = {Campañá,  Carlos and M\"{u}ser,  Martin H.},
  year = {2006},
  month = aug
}

@article{Carlson2012AFM,
  title = {Active, Programmable Elastomeric Surfaces with Tunable Adhesion for Deterministic Assembly by Transfer Printing},
  volume = {22},
  ISSN = {1616-3028},
  url = {http://dx.doi.org/10.1002/adfm.201201023},
  DOI = {10.1002/adfm.201201023},
  number = {21},
  journal = {Advanced Functional Materials},
  publisher = {Wiley},
  author = {Carlson,  Andrew and Wang,  Shuodao and Elvikis,  Paulius and Ferreira,  Placid M. and Huang,  Yonggang and Rogers,  John A.},
  year = {2012},
  month = jun,
  pages = {4476–4484}
}

@article{Ciavarella2019JMPS,
  title = {The generalized Tabor parameter for adhesive rough contacts near complete contact},
  volume = {122},
  ISSN = {0022-5096},
  url = {http://dx.doi.org/10.1016/j.jmps.2018.08.011},
  DOI = {10.1016/j.jmps.2018.08.011},
  journal = {Journal of the Mechanics and Physics of Solids},
  publisher = {Elsevier BV},
  author = {Ciavarella,  Michele and Xu,  Yang and Jackson,  Robert L.},
  year = {2019},
  month = jan,
  pages = {126–140}
}

@article{Chen2021NC,
  title = {Physics-informed learning of governing equations from scarce data},
  volume = {12},
  ISSN = {2041-1723},
  url = {http://dx.doi.org/10.1038/s41467-021-26434-1},
  DOI = {10.1038/s41467-021-26434-1},
  number = {1},
  journal = {Nature Communications},
  publisher = {Springer Science and Business Media LLC},
  author = {Chen,  Zhao and Liu,  Yang and Sun,  Hao},
  year = {2021},
  month = oct
}

@article{Ciavarella2022JA,
  title = {Maugis-Tabor parameter dependence of pull-off in viscoelastic line Hertzian contacts},
  volume = {99},
  ISSN = {1545-5823},
  url = {http://dx.doi.org/10.1080/00218464.2022.2066998},
  DOI = {10.1080/00218464.2022.2066998},
  number = {6},
  journal = {The Journal of Adhesion},
  publisher = {Informa UK Limited},
  author = {Ciavarella,  M. and Wang,  Qing-Ao and Li,  Qunyang},
  year = {2022},
  month = apr,
  pages = {972–987}
}

@article{Derjaguin1975JCIS,
  title = {Effect of contact deformations on the adhesion of particles},
  volume = {53},
  ISSN = {0021-9797},
  url = {http://dx.doi.org/10.1016/0021-9797(75)90018-1},
  DOI = {10.1016/0021-9797(75)90018-1},
  number = {2},
  journal = {Journal of Colloid and Interface Science},
  publisher = {Elsevier BV},
  author = {Derjaguin,  B.V and Muller,  V.M and Toporov,  Yu.P},
  year = {1975},
  month = nov,
  pages = {314–326}
}

@article{Dapp2012PRL,
  title = {Self-Affine Elastic Contacts: Percolation and Leakage},
  volume = {108},
  ISSN = {1079-7114},
  url = {http://dx.doi.org/10.1103/PhysRevLett.108.244301},
  DOI = {10.1103/physrevlett.108.244301},
  number = {24},
  journal = {Physical Review Letters},
  publisher = {American Physical Society (APS)},
  author = {Dapp,  Wolf B. and L\"{u}cke,  Andreas and Persson,  Bo N. J. and M\"{u}ser,  Martin H.},
  year = {2012},
  month = jun
}

@article{Greenwood1966RSPA,
  title = {Contact of nominally flat surfaces},
  volume = {295},
  ISSN = {2053-9169},
  url = {http://dx.doi.org/10.1098/rspa.1966.0242},
  DOI = {10.1098/rspa.1966.0242},
  number = {1442},
  journal = {Proceedings of the Royal Society of London. Series A. Mathematical and Physical Sciences},
  publisher = {The Royal Society},
  author = {Greenwood,  J. A. and Williamson,  J. B. P.},
  year = {1966},
  month = dec,
  pages = {300–319}
}

@article{Gao2005MM,
  title = {Mechanics of hierarchical adhesion structures of geckos},
  volume = {37},
  ISSN = {0167-6636},
  url = {http://dx.doi.org/10.1016/j.mechmat.2004.03.008},
  DOI = {10.1016/j.mechmat.2004.03.008},
  number = {2-3},
  journal = {Mechanics of Materials},
  publisher = {Elsevier BV},
  author = {Gao,  Huajian and Wang,  Xiang and Yao,  Haimin and Gorb,  Stanislav and Arzt,  Eduard},
  year = {2005},
  month = feb,
  pages = {275–285}
}

@misc{He2026ARXIV,
    doi = {10.48550/ARXIV.2603.00904},
    url = {https://arxiv.org/abs/2603.00904},
    author = {He, Roy Y. and Liang, Ying and Zhao, Hongkai and Zhong, Yimin},
    title = {From Frequency Bias to Spectral Balance: Operator-Aware Preconditioners for PINNs},
    publisher = {arXiv},
    year = {2026},
    copyright = {arXiv.org perpetual, non-exclusive license}
}

@inproceedings{Hertz1882,
  title = {On the contact of elastic solids},
  author = {Hertz, Heinrich},
  booktitle = {Journal f\"ur die reine und angewandte Mathematik},
  volume = {92},
  pages = {156--171},
  year = {1882}
}

@article{Hyun2004PRE,
  title = {Finite-element analysis of contact between elastic self-affine surfaces},
  volume = {70},
  ISSN = {1550-2376},
  url = {http://dx.doi.org/10.1103/physreve.70.026117},
  DOI = {10.1103/physreve.70.026117},
  number = {2},
  journal = {Physical Review E},
  publisher = {American Physical Society (APS)},
  author = {Hyun,  S. and Pei,  L. and Molinari,  J.-F. and Robbins,  M. O.},
  year = {2004},
  month = aug
}

@article{Johnson1971RSPA,
  title = {Surface energy and the contact of elastic solids},
  volume = {324},
  ISSN = {2053-9169},
  url = {http://dx.doi.org/10.1098/rspa.1971.0141},
  DOI = {10.1098/rspa.1971.0141},
  number = {1558},
  journal = {Proceedings of the Royal Society of London. A. Mathematical and Physical Sciences},
  publisher = {The Royal Society},
  author = {Johnson,  Kenneth Langstreth and Kendall,  Kevin and Roberts,  A. D.},
  year = {1971},
  month = sep,
  pages = {301–313}
}

@article{Kalliorinne2021FME,
  title = {Artificial Neural Network Architecture for Prediction of Contact Mechanical Response},
  volume = {6},
  ISSN = {2297-3079},
  url = {http://dx.doi.org/10.3389/fmech.2020.579825},
  DOI = {10.3389/fmech.2020.579825},
  journal = {Frontiers in Mechanical Engineering},
  publisher = {Frontiers Media SA},
  author = {Kalliorinne,  Kalle and Larsson,  Roland and Pérez-Ràfols,  Francesc and Liwicki,  Marcus and Almqvist,  Andreas},
  year = {2021},
  month = may
}

@article{Li2026JMPS,
  title = {An energy-based physics-informed neural network framework for efficient mesh-free homogenisation of architected metamaterials},
  volume = {210},
  ISSN = {0022-5096},
  url = {http://dx.doi.org/10.1016/j.jmps.2026.106542},
  DOI = {10.1016/j.jmps.2026.106542},
  journal = {Journal of the Mechanics and Physics of Solids},
  publisher = {Elsevier BV},
  author = {Li,  Haolin and Li,  Menglei and Bai,  Jinshuai and Sharif Khodaei,  Zahra and Aliabadi,  M.H.},
  year = {2026},
  month = apr,
  pages = {106542}
}

@article{Lorenz2008JPCM,
  title = {Interfacial separation between elastic solids with randomly rough surfaces: comparison of experiment with theory},
  volume = {21},
  ISSN = {1361-648X},
  url = {http://dx.doi.org/10.1088/0953-8984/21/1/015003},
  DOI = {10.1088/0953-8984/21/1/015003},
  number = {1},
  journal = {Journal of Physics: Condensed Matter},
  publisher = {IOP Publishing},
  author = {Lorenz,  B and Persson,  B N J},
  year = {2008},
  month = dec,
  pages = {015003}
}

@article{Li2024IJMS,
  title = {Regimes in the axisymmetric stiction of thin elastic plates},
  volume = {284},
  ISSN = {0020-7403},
  url = {http://dx.doi.org/10.1016/j.ijmecsci.2024.109740},
  DOI = {10.1016/j.ijmecsci.2024.109740},
  journal = {International Journal of Mechanical Sciences},
  publisher = {Elsevier BV},
  author = {Li,  Hang and Yu,  Chuanli and Dai,  Zhaohe},
  year = {2024},
  month = dec,
  pages = {109740}
}

@article{Maghami2025IJSS,
  title = {Pull-off force prediction in viscoelastic adhesive Hertzian contact by physics augmented machine learning},
  volume = {322},
  ISSN = {0020-7683},
  url = {http://dx.doi.org/10.1016/j.ijsolstr.2025.113584},
  DOI = {10.1016/j.ijsolstr.2025.113584},
  journal = {International Journal of Solids and Structures},
  publisher = {Elsevier BV},
  author = {Maghami,  Ali and Stender,  Merten and Papangelo,  Antonio},
  year = {2025},
  month = Nov,
  pages = {113584}
}

@article{Maugis1992JCIS,
  title = {Adhesion of spheres: The JKR-DMT transition using a dugdale model},
  volume = {150},
  ISSN = {0021-9797},
  url = {http://dx.doi.org/10.1016/0021-9797(92)90285-T},
  DOI = {10.1016/0021-9797(92)90285-t},
  number = {1},
  journal = {Journal of Colloid and Interface Science},
  publisher = {Elsevier BV},
  author = {Maugis,  Daniel},
  year = {1992},
  month = apr,
  pages = {243–269}
}

@article{Materzok2022S,
  title = {Gecko Adhesion on Flat and Rough Surfaces: Simulations with a Multi‐Scale Molecular Model},
  volume = {18},
  ISSN = {1613-6829},
  url = {http://dx.doi.org/10.1002/smll.202201674},
  DOI = {10.1002/smll.202201674},
  number = {35},
  journal = {Small},
  publisher = {Wiley},
  author = {Materzok,  Tobias and De Boer,  Danna and Gorb,  Stanislav and M\"{u}ller‐Plathe,  Florian},
  year = {2022},
  month = aug
}

@article{McClelland2005TL,
  title = {Contact mechanics of a flexible imprinter for photocured nanoimprint lithography},
  volume = {19},
  ISSN = {1573-2711},
  url = {http://dx.doi.org/10.1007/s11249-005-4265-6},
  DOI = {10.1007/s11249-005-4265-6},
  number = {1},
  journal = {Tribology Letters},
  publisher = {Springer Science and Business Media LLC},
  author = {McClelland,  G.M. and Rettner,  C.T. and Hart,  M.W. and Carter,  K.R. and Sanchez,  M.I. and Best,  M.E. and Terris,  B.D.},
  year = {2005},
  month = may,
  pages = {59–63}
}

@article{Menga2014RSPA,
  title = {The sliding contact of a rigid wavy surface with a viscoelastic half-space},
  volume = {470},
  ISSN = {1471-2946},
  url = {http://dx.doi.org/10.1098/rspa.2014.0392},
  DOI = {10.1098/rspa.2014.0392},
  number = {2169},
  journal = {Proceedings of the Royal Society A: Mathematical,  Physical and Engineering Sciences},
  publisher = {The Royal Society},
  author = {Menga,  N. and Putignano,  C. and Carbone,  G. and Demelio,  G. P.},
  year = {2014},
  month = sep,
  pages = {20140392}
}

@article{Monti2021PRE,
  title = {Green’s function method for dynamic contact calculations},
  volume = {103},
  ISSN = {2470-0053},
  url = {http://dx.doi.org/10.1103/PhysRevE.103.053305},
  DOI = {10.1103/physreve.103.053305},
  number = {5},
  journal = {Physical Review E},
  publisher = {American Physical Society (APS)},
  author = {Monti,  Joseph M. and Pastewka,  Lars and Robbins,  Mark O.},
  year = {2021},
  month = may
}

@article{Mueller2023PRL,
  title = {Significance of Elastic Coupling for Stresses and Leakage in Frictional Contacts},
  volume = {131},
  ISSN = {1079-7114},
  url = {http://dx.doi.org/10.1103/PhysRevLett.131.156201},
  DOI = {10.1103/physrevlett.131.156201},
  number = {15},
  journal = {Physical Review Letters},
  publisher = {American Physical Society (APS)},
  author = {M\"{u}ller,  Christian and M\"{u}ser,  Martin H. and Carbone,  Giuseppe and Menga,  Nicola},
  year = {2023},
  month = oct
}

@article{Mueser2008PRL,
  title={Rigorous Field-Theoretical Approach to the Contact Mechanics of Rough Elastic Solids},
  volume={100},
  ISSN={1079-7114},
  url={http://dx.doi.org/10.1103/PhysRevLett.100.055504},
  DOI={10.1103/physrevlett.100.055504},
  number={5},
  journal={Physical Review Letters},
  publisher={American Physical Society (APS)},
  author={Müser, Martin H.},
  year={2008},
  month=feb
}

@article{Mueser2014BJN,
  title = {Single-asperity contact mechanics with positive and negative work of adhesion: Influence of finite-range interactions and a continuum description for the squeeze-out of wetting fluids},
  volume = {5},
  ISSN = {2190-4286},
  url = {http://dx.doi.org/10.3762/bjnano.5.50},
  DOI = {10.3762/bjnano.5.50},
  journal = {Beilstein Journal of Nanotechnology},
  publisher = {Beilstein Institut},
  author = {M\"{u}ser,  Martin H},
  year = {2014},
  month = Apr,
  pages = {419–437}
}

@inbook{Papangelo2026Elsevier,
  title = {Adhesive Single and Multi-asperity Contacts},
  ISBN = {9780128035818},
  url = {http://dx.doi.org/10.1016/B978-0-443-30138-4.00009-1},
  DOI = {10.1016/b978-0-443-30138-4.00009-1},
  booktitle = {Reference Module in Materials Science and Materials Engineering},
  publisher = {Elsevier},
  author = {Papangelo,  A. and Tricarico,  M. and Maghami,  A.},
  year = {2026}
}

@article{Pastewka2013PRE,
  title = {Finite-size scaling in the interfacial stiffness of rough elastic contacts},
  volume = {87},
  ISSN = {1550-2376},
  url = {http://dx.doi.org/10.1103/PhysRevE.87.062809},
  DOI = {10.1103/physreve.87.062809},
  number = {6},
  journal = {Physical Review E},
  publisher = {American Physical Society (APS)},
  author = {Pastewka,  Lars and Prodanov,  Nikolay and Lorenz,  Boris and M\"{u}ser,  Martin H. and Robbins,  Mark O. and Persson,  Bo N. J.},
  year = {2013},
  month = jun
}

@article{Persson2001JCP,
  title = {Theory of rubber friction and contact mechanics},
  volume = {115},
  ISSN = {1089-7690},
  url = {http://dx.doi.org/10.1063/1.1388626},
  DOI = {10.1063/1.1388626},
  number = {8},
  journal = {The Journal of Chemical Physics},
  publisher = {AIP Publishing},
  author = {Persson,  B. N. J.},
  year = {2001},
  month = aug,
  pages = {3840–3861}
}

@article{Persson2002EPJE,
  title = {Adhesion between an elastic body and a randomly rough hard surface},
  volume = {8},
  ISSN = {1292-895X},
  url = {http://dx.doi.org/10.1140/epje/i2002-10025-1},
  DOI = {10.1140/epje/i2002-10025-1},
  number = {4},
  journal = {The European Physical Journal E},
  publisher = {Springer Science and Business Media LLC},
  author = {Persson,  B.N.J.},
  year = {2002},
  month = jul,
  pages = {385–401}
}

@article{Persson2004JCP,
  title = {Contact area between a viscoelastic solid and a hard,  randomly rough,  substrate},
  volume = {120},
  ISSN = {1089-7690},
  url = {http://dx.doi.org/10.1063/1.1697376},
  DOI = {10.1063/1.1697376},
  number = {18},
  journal = {The Journal of Chemical Physics},
  publisher = {AIP Publishing},
  author = {Persson,  B. N. J. and Albohr,  O. and Creton,  C. and Peveri,  V.},
  year = {2004},
  month = may,
  pages = {8779–8793}
}

@article{Persson2007PRL,
  title = {Relation between Interfacial Separation and Load: A General Theory of Contact Mechanics},
  volume = {99},
  ISSN = {1079-7114},
  url = {http://dx.doi.org/10.1103/physrevlett.99.125502},
  DOI = {10.1103/physrevlett.99.125502},
  number = {12},
  journal = {Physical Review Letters},
  publisher = {American Physical Society (APS)},
  author = {Persson,  B. N. J.},
  year = {2007},
  month = sep
}

@article{Persson2018JCP,
  title = {The dependency of adhesion and friction on electrostatic attraction},
  volume = {148},
  ISSN = {1089-7690},
  url = {http://dx.doi.org/10.1063/1.5024038},
  DOI = {10.1063/1.5024038},
  number = {14},
  journal = {The Journal of Chemical Physics},
  publisher = {AIP Publishing},
  author = {Persson,  B. N. J.},
  year = {2018},
  month = apr
}

@article{Persson2021JPCM,
  title = {General theory of electroadhesion},
  volume = {33},
  ISSN = {1361-648X},
  url = {http://dx.doi.org/10.1088/1361-648x/abe797},
  DOI = {10.1088/1361-648x/abe797},
  number = {43},
  journal = {Journal of Physics: Condensed Matter},
  publisher = {IOP Publishing},
  author = {Persson,  Bo N J},
  year = {2021},
  month = aug,
  pages = {435001}
}

@article{Persson2022TL,
  title = {Fluid Leakage in Static Rubber Seals},
  volume = {70},
  ISSN = {1573-2711},
  url = {http://dx.doi.org/10.1007/s11249-022-01573-8},
  DOI = {10.1007/s11249-022-01573-8},
  number = {2},
  journal = {Tribology Letters},
  publisher = {Springer Science and Business Media LLC},
  author = {Persson,  B. N. J.},
  year = {2022},
  month = feb
}

@article{Persson2025JCP,
  title = {Rubber wear: Experiment and theory},
  volume = {162},
  ISSN = {1089-7690},
  url = {http://dx.doi.org/10.1063/5.0248199},
  DOI = {10.1063/5.0248199},
  number = {7},
  journal = {The Journal of Chemical Physics},
  publisher = {AIP Publishing},
  author = {Persson,  B. N. J. and Xu,  R. and Miyashita,  N.},
  year = {2025},
  month = feb
}

@article{Prodanov2013TL,
  title  = {On the Contact Area and Mean Gap of Rough, Elastic Contacts:
           Dimensional Analysis, Numerical Corrections, and Reference Data},
  volume = {53},
  ISSN   = {1573-2711},
  url    = {http://dx.doi.org/10.1007/s11249-013-0282-z},
  DOI    = {10.1007/s11249-013-0282-z},
  number = {2},
  journal = {Tribology Letters},
  publisher = {Springer Science and Business Media LLC},
  author = {Prodanov, Nikolay and Dapp, Wolf B. and Müser, Martin H.},
  year   = {2013},
  month  = Dec,
  pages  = {433–448}
}

@article{Putignano2014PM,
  title = {A review of boundary elements methodologies for elastic and viscoelastic rough contact mechanics},
  volume = {17},
  ISSN = {1990-5424},
  url = {http://dx.doi.org/10.1134/S1029959914040092},
  DOI = {10.1134/s1029959914040092},
  number = {4},
  journal = {Physical Mesomechanics},
  publisher = {Pleiades Publishing Ltd},
  author = {Putignano,  C. and Carbone,  G.},
  year = {2014},
  month = oct,
  pages = {321–333}
}

@InProceedings{Rahaman2019PMLR,
  title = 	 {On the Spectral Bias of Neural Networks},
  author =       {Rahaman, Nasim and Baratin, Aristide and Arpit, Devansh and Draxler, Felix and Lin, Min and Hamprecht, Fred and Bengio, Yoshua and Courville, Aaron},
  booktitle = 	 {Proceedings of the 36th International Conference on Machine Learning},
  pages = 	 {5301--5310},
  year = 	 {2019},
  editor = 	 {Chaudhuri, Kamalika and Salakhutdinov, Ruslan},
  volume = 	 {97},
  series = 	 {Proceedings of Machine Learning Research},
  month = 	 {09--15 Jun},
  publisher =    {PMLR},
  url = 	 {https://proceedings.mlr.press/v97/rahaman19a.html},
}

@article{Raissi2019JCP,
  title = {Physics-informed neural networks: A deep learning framework for solving forward and inverse problems involving nonlinear partial differential equations},
  volume = {378},
  ISSN = {0021-9991},
  url = {http://dx.doi.org/10.1016/j.jcp.2018.10.045},
  DOI = {10.1016/j.jcp.2018.10.045},
  journal = {Journal of Computational Physics},
  publisher = {Elsevier BV},
  author = {Raissi,  M. and Perdikaris,  P. and Karniadakis,  G.E.},
  year = {2019},
  month = feb,
  pages = {686–707}
}

@article{Schmidt2009S,
  title = {Distilling Free-Form Natural Laws from Experimental Data},
  volume = {324},
  ISSN = {1095-9203},
  url = {http://dx.doi.org/10.1126/science.1165893},
  DOI = {10.1126/science.1165893},
  number = {5923},
  journal = {Science},
  publisher = {American Association for the Advancement of Science (AAAS)},
  author = {Schmidt,  Michael and Lipson,  Hod},
  year = {2009},
  month = apr,
  pages = {81–85}
}

@article{Suman2025JT,
  title = {Predictive Modeling of Real Contact Area on Rough Surfaces Using Deep Artificial Neural Network},
  volume = {147},
  ISSN = {1528-8897},
  url = {http://dx.doi.org/10.1115/1.4068057},
  DOI = {10.1115/1.4068057},
  number = {11},
  journal = {Journal of Tribology},
  publisher = {ASME International},
  author = {Suman,  Siddharth and Prajapati,  Deepak K.},
  year = {2025},
  month = mar
}

@inproceedings{Tancik2020ANIPS,
 author = {Tancik, Matthew and Srinivasan, Pratul and Mildenhall, Ben and Fridovich-Keil, Sara and Raghavan, Nithin and Singhal, Utkarsh and Ramamoorthi, Ravi and Barron, Jonathan and Ng, Ren},
 booktitle = {Advances in Neural Information Processing Systems},
 editor = {H. Larochelle and M. Ranzato and R. Hadsell and M.F. Balcan and H. Lin},
 pages = {7537--7547},
 publisher = {Curran Associates, Inc.},
 title = {Fourier Features Let Networks Learn High Frequency Functions in Low Dimensional Domains},
 url = {https://proceedings.neurips.cc/paper_files/paper/2020/file/55053683268957697aa39fba6f231c68-Paper.pdf},
 volume = {33},
 year = {2020}
}

@article{Wang2021JSC,
  title = {Understanding and Mitigating Gradient Flow Pathologies in Physics-Informed Neural Networks},
  volume = {43},
  ISSN = {1095-7197},
  url = {http://dx.doi.org/10.1137/20M1318043},
  DOI = {10.1137/20m1318043},
  number = {5},
  journal = {SIAM Journal on Scientific Computing},
  publisher = {Society for Industrial & Applied Mathematics (SIAM)},
  author = {Wang,  Sifan and Teng,  Yujun and Perdikaris,  Paris},
  year = {2021},
  month = jan,
  pages = {A3055–A3081}
}

@article{Wriggers2003CM,
  title = {Computational Contact Mechanics},
  volume = {32},
  ISSN = {1432-0924},
  url = {http://dx.doi.org/10.1007/s00466-003-0472-x},
  DOI = {10.1007/s00466-003-0472-x},
  number = {1-2},
  journal = {Computational Mechanics},
  publisher = {Springer Science and Business Media LLC},
  author = {Wriggers,  P.},
  year = {2003},
  month = sep,
  pages = {141–141}
}

@article{Xu2025TL,
  title = {Rubber Wear: History,  Mechanisms,  and Perspectives},
  volume = {73},
  ISSN = {1573-2711},
  url = {http://dx.doi.org/10.1007/s11249-025-02025-9},
  DOI = {10.1007/s11249-025-02025-9},
  number = {3},
  journal = {Tribology Letters},
  publisher = {Springer Science and Business Media LLC},
  author = {Xu,  R. and Sheng,  W. and Zhou,  F. and Persson,  B. N. J.},
  year = {2025},
  month = jun
}

@article{Xu2026TI,
  title = {Leakage at interfaces: A comprehensive study based on Persson contact mechanics theory},
  volume = {214},
  ISSN = {0301-679X},
  url = {http://dx.doi.org/10.1016/j.triboint.2025.111352},
  DOI = {10.1016/j.triboint.2025.111352},
  journal = {Tribology International},
  publisher = {Elsevier BV},
  author = {Xu,  R. and Gil,  L. and Singer,  J. and Gontard,  L. and Leverd,  W. and Persson,  B.N.J.},
  year = {2026},
  month = feb,
  pages = {111352}
}

@article{Zeng2009L,
  title = {Frictional Adhesion of Patterned Surfaces and Implications for Gecko and Biomimetic Systems},
  volume = {25},
  ISSN = {1520-5827},
  url = {http://dx.doi.org/10.1021/la900877h},
  DOI = {10.1021/la900877h},
  number = {13},
  journal = {Langmuir},
  publisher = {American Chemical Society (ACS)},
  author = {Zeng,  Hongbo and Pesika,  Noshir and Tian,  Yu and Zhao,  Boxin and Chen,  Yunfei and Tirrell,  Matthew and Turner,  Kimberly L. and Israelachvili,  Jacob N.},
  year = {2009},
  month = jun,
  pages = {7486–7495}
}

@article{Zhou2019PRB,
  title = {Solution of boundary-element problems using the fast-inertial-relaxation-engine method},
  volume = {99},
  ISSN = {2469-9969},
  url = {http://dx.doi.org/10.1103/PhysRevB.99.144103},
  DOI = {10.1103/physrevb.99.144103},
  number = {14},
  journal = {Physical Review B},
  publisher = {American Physical Society (APS)},
  author = {Zhou,  Yunong and Moseler,  Michael and M\"{u}ser,  Martin H.},
  year = {2019},
  month = apr
}

@article{Zhou2025TL,
  title = {Physics-informed neural network approach to randomly rough surface contact mechanics},
  volume = {73},
  ISSN = {1573-2711},
  url = {http://dx.doi.org/10.1007/s11249-025-02022-y},
  DOI = {10.1007/s11249-025-02022-y},
  number = {88},
  journal = {Tribology Letters},
  publisher = {Springer Science and Business Media LLC},
  author = {Zhou,  Yunong and Song,  Hengxu},
  year = {2025},
  month = jun
}

@article{Zhou2026TI-1,
  title = {Data-driven enhanced rough contact mechanics: PINN estimation of gap distribution across length scales for partial contacts},
  volume = {214},
  ISSN = {0301-679X},
  url = {http://dx.doi.org/10.1016/j.triboint.2025.111100},
  DOI = {10.1016/j.triboint.2025.111100},
  journal = {Tribology International},
  publisher = {Elsevier BV},
  author = {Zhou,  Yunong and Song,  Hengxu},
  year = {2026},
  month = feb,
  pages = {111100}
}

@article{Zhou2026TI-2,
  title = {Field-theoretical approach to estimate mean gap and gap distribution in randomly rough surface contact mechanics},
  volume = {220},
  ISSN = {0301-679X},
  url = {http://dx.doi.org/10.1016/j.triboint.2026.111894},
  DOI = {10.1016/j.triboint.2026.111894},
  journal = {Tribology International},
  publisher = {Elsevier BV},
  author = {Zhou,  Yunong and Song,  Hengxu and Zhang,  Zhichao and Xu,  Yang},
  year = {2026},
  month = aug,
  pages = {111894}
}


\end{document}